\documentclass[aps,prl,twocolumn,superscriptaddress]{revtex4-2}

\usepackage{lineno}
\usepackage{amsthm}
\usepackage{amsmath}
\usepackage{amssymb}
\usepackage{mathrsfs}
\usepackage{txfonts}
\usepackage{graphics}
\usepackage{color} 
\usepackage{enumitem}
\usepackage{bm} 

\begin{document}

\preprint{}

\title{Origins and conservation of topological polarization defects in resonant photonic-crystal 
 diffraction}


\author{Xuefan Yin}
\affiliation{Department of Electronic Science and Engineering, Kyoto University, Kyoto-Daigaku-Katsura, Nishikyo-ku, Kyoto 615-8510, Japan}
\author{Takuya Inoue}
\affiliation{Department of Electronic Science and Engineering, Kyoto University, Kyoto-Daigaku-Katsura, Nishikyo-ku, Kyoto 615-8510, Japan}
\author{Chao Peng}
\email{pengchao@pku.edu.cn}
\affiliation{State Key Laboratory of Advanced Optical Communication Systems and Networks, School of
Electronics, $\&$ Frontiers Science Center for Nano-optoelectronics,
Peking University, Beijing, 100871, China}
\affiliation{Peng Cheng Laboratory, Shenzhen 518055, China}

\author{Susumu Noda}
\email{snoda@qoe.kuee.kyoto-u.ac.jp}
\affiliation{Department of Electronic Science and
Engineering, Kyoto University, Kyoto-Daigaku-Katsura, Nishikyo-ku,
Kyoto 615-8510, Japan}


\date{\today}

\begin{abstract}
We present a continuative definition of topological charge to depict the polarization defects on any resonant diffraction orders in photonic crystal slab regardless they are radiative or evanescent. By using such a generalized definition, we investigate the origins and conservation of \textcolor{black}{integer} polarization defects across the whole Brollouin zone. We found that \textcolor{black}{these} polarization defects eventually originate from the mode degeneracy that is induced by lattice coupling as a consequence of momentum space folding, or inter-band coupling that can be either Hermitian or Non-hermitian. By counting all types of polarization defects, the total topological charge numbers in a given diffraction order is a conserved quantity across the whole Brillouin zone that is determined by lattice geometry only. 
\end{abstract}


\maketitle


  

Polarization defects \cite{nye1974dislocations,nye1983polarization,berry1998much,berry2001polarization,berry2004polarization,Flossmann_polarization_2008,cardano2013generation,thomas_observation_2015,fosel2017lines,Bliokh_2019,chen2019singularities,liu2021topological,wang2021polarization} are exotic phenomena that can happen in both real and momentum space at which one or two components that compose the light's polarization are ill-defined, corresponding to some special points on the Poincaré sphere, \textcolor{black}{such as the north or south poles which represent the circular-polarized states (CPs) \cite{nye1983lines,mokhun2002elliptic,fosel2017lines,shilei_circular} and the amplitudes singularity at the sphere center that is related to bound states in the continuum (BICs)\cite{von_neuman_uber_1929,1985Interfering,marinica2008bound,hsu_bound_2016,koshelev2020engineering,sadreev2021interference,hu2022global}}. Polarization defects bridge the underlying non-trivial physics in singular optics \cite{dennis2009singular,Soskin_2016,gbur_singular_2016} and non-Hermitian systems \cite{feng2017non,leykam2017edge,el2018non,shen2018topological,bergholtz2021exceptional} to the characteristics of far-field radiation, thus enabling many applications such as high-$Q$ cavities \cite{chen2022observation}, vortex beam generators\cite{huang2020ultrafast}, and chiral devices with circular dichroism \cite{zhang2022chiral,chen2023observation}. In particular, from the view of topological photonics \cite{lu_topological_2014,khanikaev_two-dimensional_2017,ozawa_topological_2019,wang_topological_2020}, the polarization defects in photonic crystal (PC) slabs can be characterized by quantized topological charges \cite{zhen_topological_2014,bulgakov2017bound,shilei_observation_2018,alu_experimental_2018,liuwei_global_charge,shilei_polarization_singularity_2021}. For example, the BICs carry integer topological charges, and the CPs possess half-integer charges. The topological charge provides a vivid picture to depict and manipulate the far-field radiation and paves the way to rich consequences such as merging BICs \cite{jin_topologically_2019,kang2021merging,hwang2021ultralow} and unidirectional guided resonances (UGRs) \cite{yin_observation_2020,zeng_dynamic_2021,yin2023topological}.

Although topological charges establish a valid interpretation of polarization defects, such a picture is still incompetent in clarifying several important elusiveness, mainly because it is defined on the far-field radiation, and thus, their evolution is limited inside the light cone. \textcolor{black}{For instance, it is not clear how the polarization defects originates in physics and how they evolve in the whole Brillouin zone (BZ)}. \textcolor{black}{Besides, although the conservation of topological charges in momentum space has been widely recognized, it remains elusive what the physical origin of this conservation law is, and whether or not any global conservation of topological charges exists if taking all types of polarization defects into account}. 

To address the questions mentioned above, we first propose a continuative definition of topological charge in this letter, which is not only consistent with the conventional definition in characterizing the radiative waves but also can depict the topological features in evanescent waves. As a result, such a generalized topological charge is valid in the whole BZ and can be applied to any diffraction orders of resonant modes including the non-radiative ones. \textcolor{black}{Consequently, we reveal that the symmetry-protected polarization defects root from the lattice coupling.} We also track the evolution of tunable polarization defects across the entire BZ and find they eventually come from the \textcolor{black}{inter-band} coupling effect. \textcolor{black}{Moreover, we find that the reduced BZ is not a compact manifold for polarization defects in a given diffraction order, since they can cross the BZ edge and convert to the polarization defects in adjacent orders. Nevertheless, we explain that all the topological charges originating from inter-band coupling correspond to local ``wrinkles" on polarization which always cancel out with each other, and thus the total topological charges are conserved to the lattice charge that is determined by the lattice coupling only. }

\begin{figure}[htbp] 
 \centering\label{Fig1}
 \includegraphics{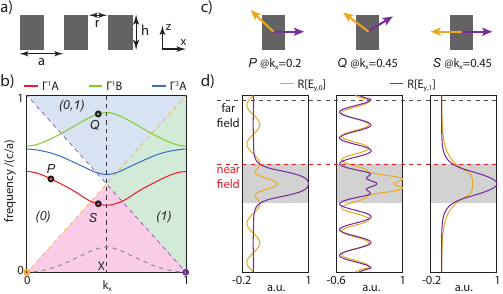} 
\caption{Resonant diffraction orders in PC slab. a, schematic of the 1D PC slab. b, energy bands of three TE modes with $r/a=0.4$ and $h/a=1.5$. Orange and purple dashed lines are light lines for 0th and 1st diffraction order, denoted by orange and purple dot, respectively. Indices in each region indicate the radiative diffraction orders. Pink region is the discrete zone beyond the light cone. c,d, momentums and profiles of 0th (orange) and 1st (purple) diffraction orders for $\mathcal{P}$, $\mathcal{Q}$ and $\mathcal{S}$ modes shown in b. Grey shading represents the slab.}
\label{Fig1} 
\end{figure}

To elaborate on our findings, we start from a 1D PC slab as shown in Fig.~\ref{Fig1}a and consider the transverse-electric (TE) modes for simplicity, while the discussion upon the 2D PC can be found in \cite{supp}. As presented in Fig.~\ref{Fig1}b, we focus on three bands in the vicinity of second-order $\Gamma$ point, denoted as $\Gamma_A^1$, $\Gamma_B^1$ and $\Gamma_A^3$, in which the postfix $A$ or $B$ represents they are either in-plane asymmetric or symmetric;  The superscript denotes the nodal numbers in the $z$-direction. 
According to Bloch's theorem, any resonant mode can be decomposed as a series of diffraction orders \cite{kogelnik1972coupled,liang_three-dimensional_2011}: $U_{y}=\sum_m E_{y, m}(z)\exp[{i(m-{k}_{x})\beta_0 x-i{k}_{y}\beta_0y]}\exp{({i\omega t})}$, with $\beta_0=2\pi/a$. Accordingly, the in-plane momentum of $m$-order diffraction is given by $\beta_{m}=(m-k_x)\beta_0$ with ${k}_{x}\beta_0$ is the offset wavevector to the \textcolor{black}{$\Gamma$ point}.

The diffraction orders can be either radiative or evanescent, determined by whether its in-plane momentum $\beta_m$ falls into the light cones as $|\beta_m|<\omega/c$ that periodically appear at any order of $\Gamma$ points. As a result, we color the regions in Fig.~\ref{Fig1}b to distinguish whether diffraction order $(m_1, m_2, ...)$ is radiative. We pick three modes $P, Q, S$ in different regions as examples. As shown in Fig.~\ref{Fig1}c, the momentum of diffraction orders $m=0$ (orange arrow) and $m=1$ (purple arrow) can be either purely in-plane or oblique, proved by their electrical fields being evanescent or oscillating (Fig.~\ref{Fig1}d).

\begin{figure}[htbp] 
 \centering \label{Fig2}
 \includegraphics{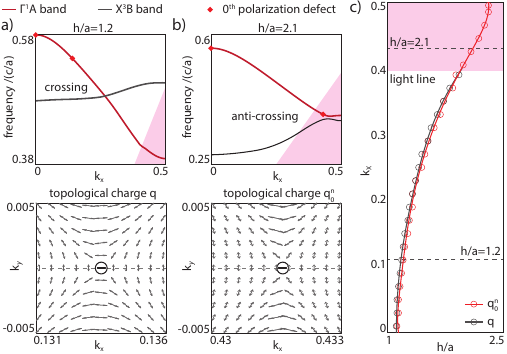} 
\caption{Near-field topological charge. a,b, band structure and polarization vector field around the defect when slab thickness $h/a=1.2$ and $2.1$. c, evolution of far-field (grey dotted line) and near-field (red dotted line) topological charge in momentum space.}
\label{Fig2} 
\end{figure}
In literature, polarization defects are usually defined in the far field (gray dashed line, Fig. \ref{Fig1}d) and thus are not valid for non-radiative diffraction orders. To address this problem, \textcolor{black}{considering the radiative waves are equivalent at any $z$ position under a gauge transformation, including the near-field position where the evanescent waves are valid, we generalize the definition of topological charges upon near-field waves (red dashed line, Fig. \ref{Fig1}d) as:} 
\begin{equation}\label{eq:1}
  q^n_m = \frac{1}{2\pi} \oint_C d\mathbf{k} \cdot \nabla_{\mathbf{k}} \theta^n_m(\mathbf{k})
\end{equation}
Here, $\theta^n_m(\mathbf{k})$ is the major angle of polarization vector defined by near-field amplitude  $(c^n_{m,x},c^n_{m,y})$ of $m$-order diffraction. Obviously, $q_m$ is valid for any diffraction orders regardless they are radiative or evanescent, and the conventional topological charge $q$ in the far-field polarization is a subset of $q_m^n$.

The above continuative definition of $q_m^n$ allows us to learn how polarization defects cross the light line and evolve in the whole BZ. To show this fact, we take the $\Gamma_A^1$ band as an example (Fig. \ref{Fig2}), in which only 0th diffraction \textcolor{black}{is radiative in white region and becomes evanescent in pink region}.
\textcolor{black}{Two polarization defects with integer charges (denoted as integer defect for short) can be found on $\Gamma_A^1$ band (red dot, Fig. \ref{Fig2}a): One is fixed at the $\Gamma$ point corresponding to symmetry-protected BIC; another one locates at $k_x=0.133$ when $h/a=1.2$, corresponding to a tunable BIC and evolves along the $k_x$ axis when parameter varies.} The latter carries a conventional far-field topological charge of $q=-1$ (lower panel, Fig. \ref{Fig2}a). By increasing the slab thickness, $q$ evolves toward a large $k_x$  but becomes invalid when it reaches the light line. Instead, the full evolution of such an integer defect can be captured by the generalized topological charge $q_0^n$. A near-field integer defect is found below the light line at $k_x=0.431$ when $h/a=2.1$ (red dot, Fig. \ref{Fig2}b), carrying a $q_0^n=-1$ (lower panel, Fig. \ref{Fig2}b). We plot the trajectories of both $q$ and $q_0^n$ under continuously varying thickness ( Fig. \ref{Fig2}c), showing that they overlap in the radiative region but $q_0^n$ smoothly cross the light cone and evolve to the BZ edge. 

\begin{figure}[htbp] 
 \centering \label{Fig3}
 \includegraphics{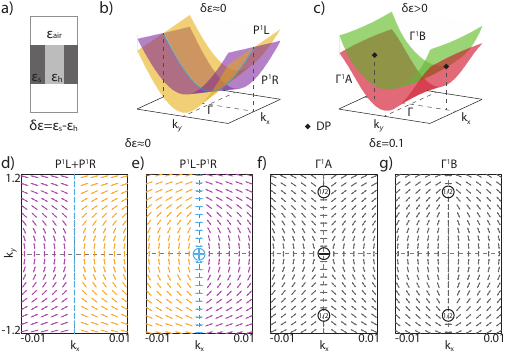} 
\caption{Generation of integer defects at $\Gamma$ point. a, schematic of unit cell of 1D PC slab. b,c, band structures when $\delta_\varepsilon\approx 0$ and $\delta_\varepsilon> 0$. d,e (f,g), polarization vector fields for $P_L^1$ and $P_R^3$ ($\Gamma_A^1$ and $X_B^3$) when $\delta_\varepsilon\approx 0$ ($\delta_\varepsilon=0.1$)}
\label{Fig3} 
\end{figure}

\textcolor{black}{It is interesting and important to ask about the origins of integer defects. We first consider the defect right at the $\Gamma$ point which is generated from the coupling between two counter-propagating modes \cite{koshelev2020engineering,hu2022global}, described by a two-level model:}
\begin{equation}
    \label{eq:21}
\hat{\mathcal{H}}_0=\omega_0+\eta k_y^2+(\Delta_x+\Delta_y k_y^2)\sigma_x+\xi k_x\sigma_y
\end{equation}
\textcolor{black}{where $\omega_0$ is the degenerate frequency of propagating modes at $\Gamma$ point, and $\eta,~\xi$ are coefficients associating with dispersion. Parameters $\Delta_x$ and $\Delta_y$ describe the lattice coupling and thus depend on $\delta_\varepsilon=\varepsilon_s-\varepsilon_h$. For a homogeneous slab ($\delta_\varepsilon=0$, Fig. 3a), both $\Delta_x$ and $\Delta_y$ are zero, giving rise to two degenerate eigenvectors $[1,0]^T$ and $[0,1]^T$ at $\Gamma$ point ($k_x=0$), representing the left-propagating mode (denoted as $P^1_L$) and right-propagating mode ($P^1_R$), respectively. As shown in Fig. \ref{Fig3}b, their bands cross at $\Gamma$ point (blue line), where the eigenstates can be any linear combination of original counter-propagating waves. In this case no diffraction orders can be defined.}

\textcolor{black}{To investigate the 0th diffraction, we consider an infinitesimal $\delta_\varepsilon$ which doesn't lift the degeneracy along $k_y$ axis but create an infinitesimal 0th diffraction order. Similar to the physics of spontaneous symmetry breaking \cite{abud1983geometry,beekman2019introduction}, the linear combination of eigenstates along $k_y$ axis can be determined according to the lattice geometry: $P^1_L\pm P^1_R=[1,\pm1]^T$. Since the continuative definition of topological charge can track the polarization defects beyond the light cone, we plot the polarization vector fields of $P^1_R\pm P^1_L$ modes, shown in Fig. \ref{Fig3}d and e. The total topological charge for mode $P^1_R+P^1_L$ is zero ($q_{0,+}^n=0$) while a nonzero integer charge can be found for mode $P^1_R-P^1_L$ ($q_{0,-}^n=1$). This nonzero charge doesn't correspond to a physical polarization defect but a consequence of spontaneous symmetry breaking of two-fold degeneracy under the gauge of lattice geometry at $\Gamma$ point. We denote it as ``lattice charge" \cite{supp}.}

\textcolor{black}{Next, we consider a nonzero $\delta_\varepsilon$ to lift the degeneracy ($\Delta_{x,y}\neq 0$), and modes $P^1_L\pm P^1_R$ lift to separated $\Gamma^1_{A,B}$ modes, respectively. An integer topological charge of $q_0^n=-1$ can be found in $\Gamma^1_A$ mode, corresponding to the symmetry-protected BIC. Besides, two Dirac points (DPs) can be found at $(k_x=0,k_y=\pm\sqrt{\Delta_x/\Delta_y})$ (black dots, Fig. \ref{Fig3}c), each carrying a half-integer charges of $q_0^n=1/2$ (denoted as half defect for short) on the $k_y$ axis for both $\Gamma^1_A$ and $\Gamma^1_B$ modes. As a result, we find that the total topological charges for $\Gamma^1_{A,B}$ modes actually conserve to the lattice charge: $q_{0,+}^n=q_{0,A}^n=0$ and $q_{0,-}^n=q_{0,B}^n=1$, shown in Fig. \ref{Fig3}f and g. This conservation reveal that the integer defect fixed at $\Gamma$ point origins from the lattice charge created by spontaneous symmetry breaking. We argue that such conserved charge number is determined by the lattice geometry only but not related to the unit cell geometry.} \textcolor{black}{When applying other unit cell geometry, different polarization defects configuration can be found, but still conserved to the lattice charge \cite{supp}.}

\begin{figure}[htbp] 
 \centering \label{Fig4}
 \includegraphics{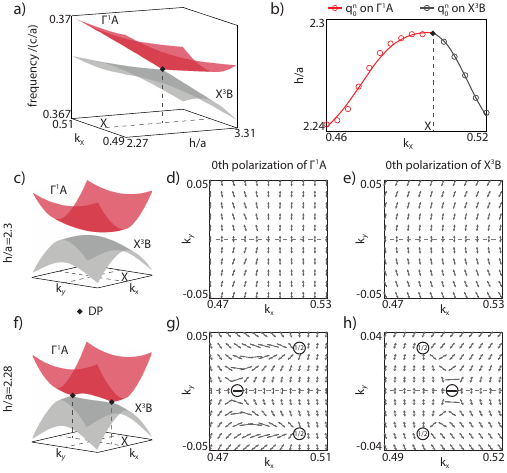} 
\caption{Generation of tunable integer defects: Hermitian case. a, coupling scenario of $\Gamma_A^1$ and $X_B^3$ modes. b, evolution of 0th polarization defects on $\Gamma_A^1$ (red dotted line) and $X_B^3$ (gray dotted line). c,d,e (f,g,h), band structures, polarization vector field for $\Gamma_A^1$ and $X_B^3$ when $h/a=2.3 
 (h/a=2.28)$.}
\label{Fig4} 
\end{figure}
Further, we investigate the \textcolor{black}{origin of tunable} integer charge. \textcolor{black}{As presented in Fig. \ref{Fig2}a, when slab thickness is relatively thin, the quality factor of $X^3_B$ mode is very low, and thus it simply cross with $\Gamma^1_A$ band without strong coupling. When slab thickness becomes thicker, two bands clearly exhibit an anti-crossing feature (Fig. \ref{Fig2}b),} suggesting such a movable integer defect is raised by the interband coupling between them. \textcolor{black}{This interband coupling scenario in 2D parameter space $(k_x,k_y=0,h)$ near the $X$ point is illustrated in Fig. \ref{Fig4}a, where a DP is found at $(k_{x0}=0.5,h_0=2.29a)$ (black dot).} \textcolor{black}{Right at the DP, all the diffraction orders are ill-defined because of the degeneracy, indicating that the eigenstates can be mixed to any ratio to produce any amplitudes of diffraction, including zero that corresponds to the integer defect.} \textcolor{black}{In other words, we conclude that the integer defects on both $\Gamma_A^1$ and $X_B^3$ bands are spawned from the DP here, confirmed by the trajectories shown in Fig. \ref{Fig4}b}.

\textcolor{black}{We verify that the total topological charge is conserved during the interband coupling process. Taking 0th diffraction as an example,} in the case without coupling ($h/a=2.3$, Fig. \ref{Fig4}c), we got $q^n_0=0$ for both $\Gamma_A^1$ and $X_B^3$ bands (Fig. \ref{Fig4}d and e). \textcolor{black}{When slab thickness decreases, two bands touch at the $X$ point that give rise to a DP at $(k_x=0.5,k_y=0)$. By further decreasing the thickness to $h/a=2.28$, the $X$-point DP splits to two DPs at $(k_x=0.5,k_y=\pm 0.03)$ (black dots, Fig. \ref{Fig4}f), each carrying a half-charge of $q^n_0=1/2$ (black dots, Fig. \ref{Fig4}g and h).} At the same time, two integer defects $q^n_0=-1$ appear on both bands (Fig. \ref{Fig4}g and h). The total 0th topological charges are conserved for both bands, followed as $q^n_{all}=1/2+1/2-1=0$ which is the same as the uncoupled case. This local conservation law indicates that to compensate for the nonzero charges created by the DPs, some integer or half-integer polarization defects must be generated simultaneously.  

\textcolor{black}{After spawned from the DP at the BZ edge, integer defects on both $\Gamma_A^1$ and $X_B^3$ bands robustly evolve in momentum space, departing away from each other due to frequency shifting when slab thickness decreasing}. \textcolor{black}{The defect on $X_B^3$ band moves away from the first BZ, while the defect on $\Gamma_A^1$ band gradually approaches the $\Gamma$ point and eventually turns into a BIC after falling into the light cone.} Noteworthy is that the 0th diffraction at $k_x$ point is actually equivalent to the adjacent -1st order diffraction at $k_x-1$ point according to the \textcolor{black}{same} momentum. If the 0th polarization defect leaves the first BZ by crossing the $X$ point, it turns out to be a -1st polarization defect that enters the first BZ from the $-X$ point. This fact indicates that the polarization defects on different diffraction orders can convert into each other by crossing the BZ edge. 

\begin{figure}[htbp] 
 \centering 
 \includegraphics{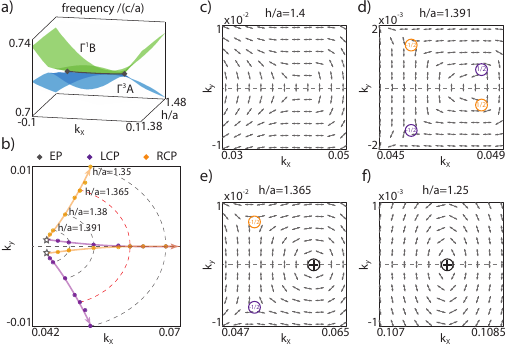} 
\caption{Generation of tunable integer defects: non-Hermitian case. a, coupling scenario of $\Gamma_A^3$ and $\Gamma_B^1$ modes. b, evolution of CPs on $\Gamma_A^3$ band when slab thickness varies. c,d,e,f, polarization vector field of $\Gamma_A^3$ mode when $h/a=1.4$, $1.391$, $1.365$ and $1.25$.}
\label{Fig5}
\end{figure}
\textcolor{black}{By applying a similar argument, we found that the polarization defect can also emerge from non-Hermitian inter-band coupling around the BZ center.} Taking the structure shown in Fig. \ref{Fig1} for instance, at a certain slab thickness,  $\Gamma_B^1$ and $\Gamma_A^3$ bands get coupled to each other around the $\Gamma$ point. As shown in Fig. \ref{Fig5}a, a pair of exceptional points (EPs) \cite{kato2013perturbation,berry2004physics,heiss2012physics,zhen2015spawning,doppler2016dynamically,zhou_observation_2018,miri2019exceptional} is found at $(k_x=\pm 0.05,h/a=1.43)$ (grey dots), \textcolor{black}{featuring the non-Hermitian interband coupling, which is different from an isolated DP in Hermitian case.} \textcolor{black}{Nevertheless, polarization defects can similarly be created owing to the interband coupling, but they show a more complex evolution without the DP. The full charge evolution can be found in \cite{supp}. Here we plot the trajectories of topological charges in $\Gamma_A^3$ band for an example, shown in Fig. \ref{Fig5}b. In a word, when decreasing the slab thickness, the integer defect acts as the merging point of two CPs carrying the same-signed half charges of $q_0^n=1/2$, corresponding to a Friedrich–Wintgen (FW) BIC \cite{1985Interfering}.} \textcolor{black}{Once generated, the integer defect robustly evolves along the $k_x$ axis until it enters the first diffraction zone \cite{supp}. If all possible half-charges are counted together, the total topological charges in the region of interest (ROI) are still conserved as $q^n_{0,all}=2\times(1/2-1/2)=0$, which is similar with the Hermitian case.}

Combining the above discussion, we conclude the global conservation of topological charges, namely to check if there exists a topological invariant to represent the total topological charge numbers in a given diffraction order across the whole BZ. 
\textcolor{black}{Basically, all the polarization defects can be considered to generate from the band coupling, which has two origins: one is the lattice charges originated from spontaneous symmetry breaking between two-fold degenerate propagating modes induced by lattice coupling; the other is the inter-band coupling between two accidental crossed bloch bands.} However, different from other topological invariant such as the Chern number \cite{hatsugai1993chern}, polarization defects belong to the topological feature of diffraction orders but not the eigenstate itself. Therefore, the reduced BZ is not a compact manifold for polarization defects in a given diffraction order but the entire BZ should be considered. Since the polarization defects raised by inter-band coupling are actually local wrinkles on polarization due to Stokes' theorem, they are always conserved to zero if taking all the generated integer and half defects into account. As a result, the global topological charge across the whole BZ is only determined by lattice geometry, as a consequence of the folding of momentum space.

\textcolor{black}{At last,} we point out that, in addition to the BICs, the picture of polarization defects we established above can be applied to other radiation singularities such as UGRs which can be interpreted as the polarization defects only on one side \cite{yin2023topological}. Actually, the polarization defect are a universal phenomenon and we can create them by folding the momentum space to form more band crossings. \textcolor{black}{The codimension of polarization defects can be affected by the system symmetry as well as the number of radiative channels. For system with multiple radiative diffraction orders, extra degrees of freedom are necessary to make polarization defects robust in parameter space. More discussion and examples are presented in \cite{supp}.}

To summarize, we present a continuative definition of topological charge in the near-field waves in this work, which allows us to track and discuss the origins and conservation of polarization defects on any resonant diffraction orders across the whole BZ. We found that \textcolor{black}{polarization defects are not a unique feature of non-Hermitian system, but origin from the band coupling associating with more fundamental physics in both Hermitian and non-Hermitian system:} they eventually come from the spontaneous symmetry breaking of degenerate propagating waves as a result of folded momentum space, \textcolor{black}{or inter-band coupling induced by accidental band crossing. Moreover, we revealed that,} by taking all possible topological charges into account, the total topological charge number in a given diffraction order across the whole BZ is conserved to a quantity that is determined by lattice geometry only.


\bibliography{reference.bib}

\begin{thebibliography}{64}%
\makeatletter
\providecommand \@ifxundefined [1]{%
 \@ifx{#1\undefined}
}%
\providecommand \@ifnum [1]{%
 \ifnum #1\expandafter \@firstoftwo
 \else \expandafter \@secondoftwo
 \fi
}%
\providecommand \@ifx [1]{%
 \ifx #1\expandafter \@firstoftwo
 \else \expandafter \@secondoftwo
 \fi
}%
\providecommand \natexlab [1]{#1}%
\providecommand \enquote  [1]{``#1''}%
\providecommand \bibnamefont  [1]{#1}%
\providecommand \bibfnamefont [1]{#1}%
\providecommand \citenamefont [1]{#1}%
\providecommand \href@noop [0]{\@secondoftwo}%
\providecommand \href [0]{\begingroup \@sanitize@url \@href}%
\providecommand \@href[1]{\@@startlink{#1}\@@href}%
\providecommand \@@href[1]{\endgroup#1\@@endlink}%
\providecommand \@sanitize@url [0]{\catcode `\\12\catcode `\$12\catcode
  `\&12\catcode `\#12\catcode `\^12\catcode `\_12\catcode `\%12\relax}%
\providecommand \@@startlink[1]{}%
\providecommand \@@endlink[0]{}%
\providecommand \url  [0]{\begingroup\@sanitize@url \@url }%
\providecommand \@url [1]{\endgroup\@href {#1}{\urlprefix }}%
\providecommand \urlprefix  [0]{URL }%
\providecommand \Eprint [0]{\href }%
\providecommand \doibase [0]{https://doi.org/}%
\providecommand \selectlanguage [0]{\@gobble}%
\providecommand \bibinfo  [0]{\@secondoftwo}%
\providecommand \bibfield  [0]{\@secondoftwo}%
\providecommand \translation [1]{[#1]}%
\providecommand \BibitemOpen [0]{}%
\providecommand \bibitemStop [0]{}%
\providecommand \bibitemNoStop [0]{.\EOS\space}%
\providecommand \EOS [0]{\spacefactor3000\relax}%
\providecommand \BibitemShut  [1]{\csname bibitem#1\endcsname}%
\let\auto@bib@innerbib\@empty
\bibitem [{\citenamefont {Nye}\ and\ \citenamefont
  {Berry}(1974)}]{nye1974dislocations}%
  \BibitemOpen
  \bibfield  {author} {\bibinfo {author} {\bibfnamefont {J.~F.}\ \bibnamefont
  {Nye}}\ and\ \bibinfo {author} {\bibfnamefont {M.~V.}\ \bibnamefont
  {Berry}},\ }\bibinfo {title} {Dislocations in wave trains},\ in\ \href
  {https://doi.org/10.1142/9789813221215_0001} {\emph {\bibinfo {booktitle} {A
  Half-Century of Physical Asymptotics and Other Diversions}}}\ (\bibinfo
  {year} {1974})\ Chap.\ \bibinfo {chapter} {1.1}, pp.\ \bibinfo {pages}
  {6--31}\BibitemShut {NoStop}%
\bibitem [{\citenamefont {Nye}(1983{\natexlab{a}})}]{nye1983polarization}%
  \BibitemOpen
  \bibfield  {author} {\bibinfo {author} {\bibfnamefont {J.~F.}\ \bibnamefont
  {Nye}},\ }\bibfield  {title} {\bibinfo {title} {Polarization effects in the
  diffraction of electromagnetic waves: the role of disclinations},\
  }\href@noop {} {\bibfield  {journal} {\bibinfo  {journal} {Proc. R. Soc.
  Lond., A Math. phys. sci.}\ }\textbf {\bibinfo {volume} {387}},\ \bibinfo
  {pages} {105} (\bibinfo {year} {1983}{\natexlab{a}})}\BibitemShut {NoStop}%
\bibitem [{\citenamefont {Berry}(1998)}]{berry1998much}%
  \BibitemOpen
  \bibfield  {author} {\bibinfo {author} {\bibfnamefont {M.~V.}\ \bibnamefont
  {Berry}},\ }\bibfield  {title} {\bibinfo {title} {Much ado about nothing:
  optical distortion lines (phase singularities, zeros, and vortices)},\ }in\
  \href@noop {} {\emph {\bibinfo {booktitle} {International Conference on
  Singular Optics}}},\ Vol.\ \bibinfo {volume} {3487}\ (\bibinfo {year}
  {1998})\ pp.\ \bibinfo {pages} {1--5}\BibitemShut {NoStop}%
\bibitem [{\citenamefont {Berry}\ and\ \citenamefont
  {Dennis}(2001)}]{berry2001polarization}%
  \BibitemOpen
  \bibfield  {author} {\bibinfo {author} {\bibfnamefont {M.}~\bibnamefont
  {Berry}}\ and\ \bibinfo {author} {\bibfnamefont {M.}~\bibnamefont {Dennis}},\
  }\bibfield  {title} {\bibinfo {title} {Polarization singularities in
  isotropic random vector waves},\ }\href@noop {} {\bibfield  {journal}
  {\bibinfo  {journal} {Proc. R. Soc. Lond., A Math. phys. sci.}\ }\textbf
  {\bibinfo {volume} {457}},\ \bibinfo {pages} {141} (\bibinfo {year}
  {2001})}\BibitemShut {NoStop}%
\bibitem [{\citenamefont {Berry}\ \emph {et~al.}(2004)\citenamefont {Berry},
  \citenamefont {Dennis},\ and\ \citenamefont {Lee}}]{berry2004polarization}%
  \BibitemOpen
  \bibfield  {author} {\bibinfo {author} {\bibfnamefont {M.}~\bibnamefont
  {Berry}}, \bibinfo {author} {\bibfnamefont {M.}~\bibnamefont {Dennis}},\ and\
  \bibinfo {author} {\bibfnamefont {R.}~\bibnamefont {Lee}},\ }\bibfield
  {title} {\bibinfo {title} {Polarization singularities in the clear sky},\
  }\href@noop {} {\bibfield  {journal} {\bibinfo  {journal} {New J. Phys.}\
  }\textbf {\bibinfo {volume} {6}},\ \bibinfo {pages} {162} (\bibinfo {year}
  {2004})}\BibitemShut {NoStop}%
\bibitem [{\citenamefont {Flossmann}\ \emph {et~al.}(2008)\citenamefont
  {Flossmann}, \citenamefont {O`Holleran}, \citenamefont {Dennis},\ and\
  \citenamefont {Padgett}}]{Flossmann_polarization_2008}%
  \BibitemOpen
  \bibfield  {author} {\bibinfo {author} {\bibfnamefont {F.}~\bibnamefont
  {Flossmann}}, \bibinfo {author} {\bibfnamefont {K.}~\bibnamefont
  {O`Holleran}}, \bibinfo {author} {\bibfnamefont {M.~R.}\ \bibnamefont
  {Dennis}},\ and\ \bibinfo {author} {\bibfnamefont {M.~J.}\ \bibnamefont
  {Padgett}},\ }\bibfield  {title} {\bibinfo {title} {Polarization
  singularities in 2d and 3d speckle fields},\ }\href
  {https://doi.org/10.1103/PhysRevLett.100.203902} {\bibfield  {journal}
  {\bibinfo  {journal} {Phys. Rev. Lett.}\ }\textbf {\bibinfo {volume} {100}},\
  \bibinfo {pages} {203902} (\bibinfo {year} {2008})}\BibitemShut {NoStop}%
\bibitem [{\citenamefont {Cardano}\ \emph {et~al.}(2013)\citenamefont
  {Cardano}, \citenamefont {Karimi}, \citenamefont {Marrucci}, \citenamefont
  {de~Lisio},\ and\ \citenamefont {Santamato}}]{cardano2013generation}%
  \BibitemOpen
  \bibfield  {author} {\bibinfo {author} {\bibfnamefont {F.}~\bibnamefont
  {Cardano}}, \bibinfo {author} {\bibfnamefont {E.}~\bibnamefont {Karimi}},
  \bibinfo {author} {\bibfnamefont {L.}~\bibnamefont {Marrucci}}, \bibinfo
  {author} {\bibfnamefont {C.}~\bibnamefont {de~Lisio}},\ and\ \bibinfo
  {author} {\bibfnamefont {E.}~\bibnamefont {Santamato}},\ }\bibfield  {title}
  {\bibinfo {title} {Generation and dynamics of optical beams with polarization
  singularities},\ }\href@noop {} {\bibfield  {journal} {\bibinfo  {journal}
  {Opt. Express.}\ }\textbf {\bibinfo {volume} {21}},\ \bibinfo {pages} {8815}
  (\bibinfo {year} {2013})}\BibitemShut {NoStop}%
\bibitem [{\citenamefont {Bauer}\ \emph {et~al.}(2015)\citenamefont {Bauer},
  \citenamefont {Banzer}, \citenamefont {Karimi}, \citenamefont {Orlov},
  \citenamefont {Rubano}, \citenamefont {Marrucci}, \citenamefont {Santamato},
  \citenamefont {Boyd},\ and\ \citenamefont
  {Leuchs}}]{thomas_observation_2015}%
  \BibitemOpen
  \bibfield  {author} {\bibinfo {author} {\bibfnamefont {T.}~\bibnamefont
  {Bauer}}, \bibinfo {author} {\bibfnamefont {P.}~\bibnamefont {Banzer}},
  \bibinfo {author} {\bibfnamefont {E.}~\bibnamefont {Karimi}}, \bibinfo
  {author} {\bibfnamefont {S.}~\bibnamefont {Orlov}}, \bibinfo {author}
  {\bibfnamefont {A.}~\bibnamefont {Rubano}}, \bibinfo {author} {\bibfnamefont
  {L.}~\bibnamefont {Marrucci}}, \bibinfo {author} {\bibfnamefont
  {E.}~\bibnamefont {Santamato}}, \bibinfo {author} {\bibfnamefont {R.~W.}\
  \bibnamefont {Boyd}},\ and\ \bibinfo {author} {\bibfnamefont
  {G.}~\bibnamefont {Leuchs}},\ }\bibfield  {title} {\bibinfo {title}
  {Observation of optical polarization m{\"o}bius strips},\ }\href
  {https://doi.org/10.1126/science.1260635} {\bibfield  {journal} {\bibinfo
  {journal} {Science}\ }\textbf {\bibinfo {volume} {347}},\ \bibinfo {pages}
  {964} (\bibinfo {year} {2015})}\BibitemShut {NoStop}%
\bibitem [{\citenamefont {F{\"o}sel}\ \emph {et~al.}(2017)\citenamefont
  {F{\"o}sel}, \citenamefont {Peano},\ and\ \citenamefont
  {Marquardt}}]{fosel2017lines}%
  \BibitemOpen
  \bibfield  {author} {\bibinfo {author} {\bibfnamefont {T.}~\bibnamefont
  {F{\"o}sel}}, \bibinfo {author} {\bibfnamefont {V.}~\bibnamefont {Peano}},\
  and\ \bibinfo {author} {\bibfnamefont {F.}~\bibnamefont {Marquardt}},\
  }\bibfield  {title} {\bibinfo {title} {L lines, c points and chern numbers:
  understanding band structure topology using polarization fields},\
  }\href@noop {} {\bibfield  {journal} {\bibinfo  {journal} {New J. Phys.}\
  }\textbf {\bibinfo {volume} {19}},\ \bibinfo {pages} {115013} (\bibinfo
  {year} {2017})}\BibitemShut {NoStop}%
\bibitem [{\citenamefont {Bliokh}\ \emph {et~al.}(2019)\citenamefont {Bliokh},
  \citenamefont {Alonso},\ and\ \citenamefont {Dennis}}]{Bliokh_2019}%
  \BibitemOpen
  \bibfield  {author} {\bibinfo {author} {\bibfnamefont {K.~Y.}\ \bibnamefont
  {Bliokh}}, \bibinfo {author} {\bibfnamefont {M.~A.}\ \bibnamefont {Alonso}},\
  and\ \bibinfo {author} {\bibfnamefont {M.~R.}\ \bibnamefont {Dennis}},\
  }\bibfield  {title} {\bibinfo {title} {Geometric phases in 2d and 3d
  polarized fields: geometrical, dynamical, and topological aspects},\ }\href
  {https://doi.org/10.1088/1361-6633/ab4415} {\bibfield  {journal} {\bibinfo
  {journal} {Rep. Prog. Phys.}\ }\textbf {\bibinfo {volume} {82}},\ \bibinfo
  {pages} {122401} (\bibinfo {year} {2019})}\BibitemShut {NoStop}%
\bibitem [{\citenamefont {Chen}\ \emph {et~al.}(2019)\citenamefont {Chen},
  \citenamefont {Chen},\ and\ \citenamefont {Liu}}]{chen2019singularities}%
  \BibitemOpen
  \bibfield  {author} {\bibinfo {author} {\bibfnamefont {W.}~\bibnamefont
  {Chen}}, \bibinfo {author} {\bibfnamefont {Y.}~\bibnamefont {Chen}},\ and\
  \bibinfo {author} {\bibfnamefont {W.}~\bibnamefont {Liu}},\ }\bibfield
  {title} {\bibinfo {title} {Singularities and poincar{\'e} indices of
  electromagnetic multipoles},\ }\href@noop {} {\bibfield  {journal} {\bibinfo
  {journal} {Phys. Rev. Lett.}\ }\textbf {\bibinfo {volume} {122}},\ \bibinfo
  {pages} {153907} (\bibinfo {year} {2019})}\BibitemShut {NoStop}%
\bibitem [{\citenamefont {Liu}\ \emph {et~al.}(2021)\citenamefont {Liu},
  \citenamefont {Liu}, \citenamefont {Shi},\ and\ \citenamefont
  {Kivshar}}]{liu2021topological}%
  \BibitemOpen
  \bibfield  {author} {\bibinfo {author} {\bibfnamefont {W.}~\bibnamefont
  {Liu}}, \bibinfo {author} {\bibfnamefont {W.}~\bibnamefont {Liu}}, \bibinfo
  {author} {\bibfnamefont {L.}~\bibnamefont {Shi}},\ and\ \bibinfo {author}
  {\bibfnamefont {Y.}~\bibnamefont {Kivshar}},\ }\bibfield  {title} {\bibinfo
  {title} {Topological polarization singularities in metaphotonics},\
  }\href@noop {} {\bibfield  {journal} {\bibinfo  {journal} {Nanophotonics}\
  }\textbf {\bibinfo {volume} {10}},\ \bibinfo {pages} {1469} (\bibinfo {year}
  {2021})}\BibitemShut {NoStop}%
\bibitem [{\citenamefont {Wang}\ \emph {et~al.}(2021)\citenamefont {Wang},
  \citenamefont {Tu}, \citenamefont {Li},\ and\ \citenamefont
  {Wang}}]{wang2021polarization}%
  \BibitemOpen
  \bibfield  {author} {\bibinfo {author} {\bibfnamefont {Q.}~\bibnamefont
  {Wang}}, \bibinfo {author} {\bibfnamefont {C.-H.}\ \bibnamefont {Tu}},
  \bibinfo {author} {\bibfnamefont {Y.-N.}\ \bibnamefont {Li}},\ and\ \bibinfo
  {author} {\bibfnamefont {H.-T.}\ \bibnamefont {Wang}},\ }\bibfield  {title}
  {\bibinfo {title} {Polarization singularities: Progress, fundamental physics,
  and prospects},\ }\href@noop {} {\bibfield  {journal} {\bibinfo  {journal}
  {Apl Photonics}\ }\textbf {\bibinfo {volume} {6}} (\bibinfo {year}
  {2021})}\BibitemShut {NoStop}%
\bibitem [{\citenamefont {Nye}(1983{\natexlab{b}})}]{nye1983lines}%
  \BibitemOpen
  \bibfield  {author} {\bibinfo {author} {\bibfnamefont {J.~F.}\ \bibnamefont
  {Nye}},\ }\bibfield  {title} {\bibinfo {title} {Lines of circular
  polarization in electromagnetic wave fields},\ }\href@noop {} {\bibfield
  {journal} {\bibinfo  {journal} {Proc. R. Soc. Lond., A Math. phys. sci.}\
  }\textbf {\bibinfo {volume} {389}},\ \bibinfo {pages} {279} (\bibinfo {year}
  {1983}{\natexlab{b}})}\BibitemShut {NoStop}%
\bibitem [{\citenamefont {Mokhun}\ \emph {et~al.}(2002)\citenamefont {Mokhun},
  \citenamefont {Soskin},\ and\ \citenamefont {Freund}}]{mokhun2002elliptic}%
  \BibitemOpen
  \bibfield  {author} {\bibinfo {author} {\bibfnamefont {A.}~\bibnamefont
  {Mokhun}}, \bibinfo {author} {\bibfnamefont {M.}~\bibnamefont {Soskin}},\
  and\ \bibinfo {author} {\bibfnamefont {I.}~\bibnamefont {Freund}},\
  }\bibfield  {title} {\bibinfo {title} {Elliptic critical points: C-points,
  a-lines, and the sign rule},\ }\href@noop {} {\bibfield  {journal} {\bibinfo
  {journal} {Opt. Lett.}\ }\textbf {\bibinfo {volume} {27}},\ \bibinfo {pages}
  {995} (\bibinfo {year} {2002})}\BibitemShut {NoStop}%
\bibitem [{\citenamefont {Liu}\ \emph {et~al.}(2019)\citenamefont {Liu},
  \citenamefont {Wang}, \citenamefont {Zhang}, \citenamefont {Wang},
  \citenamefont {Zhao}, \citenamefont {Guan}, \citenamefont {Liu},
  \citenamefont {Shi},\ and\ \citenamefont {Zi}}]{shilei_circular}%
  \BibitemOpen
  \bibfield  {author} {\bibinfo {author} {\bibfnamefont {W.}~\bibnamefont
  {Liu}}, \bibinfo {author} {\bibfnamefont {B.}~\bibnamefont {Wang}}, \bibinfo
  {author} {\bibfnamefont {Y.}~\bibnamefont {Zhang}}, \bibinfo {author}
  {\bibfnamefont {J.}~\bibnamefont {Wang}}, \bibinfo {author} {\bibfnamefont
  {M.}~\bibnamefont {Zhao}}, \bibinfo {author} {\bibfnamefont {F.}~\bibnamefont
  {Guan}}, \bibinfo {author} {\bibfnamefont {X.}~\bibnamefont {Liu}}, \bibinfo
  {author} {\bibfnamefont {L.}~\bibnamefont {Shi}},\ and\ \bibinfo {author}
  {\bibfnamefont {J.}~\bibnamefont {Zi}},\ }\bibfield  {title} {\bibinfo
  {title} {Circularly polarized states spawning from bound states in the
  continuum},\ }\href {https://doi.org/10.1103/PhysRevLett.123.116104}
  {\bibfield  {journal} {\bibinfo  {journal} {Phys. Rev. Lett.}\ }\textbf
  {\bibinfo {volume} {123}},\ \bibinfo {pages} {116104} (\bibinfo {year}
  {2019})}\BibitemShut {NoStop}%
\bibitem [{\citenamefont {von Neuman}\ and\ \citenamefont
  {Wigner}(1929)}]{von_neuman_uber_1929}%
  \BibitemOpen
  \bibfield  {author} {\bibinfo {author} {\bibfnamefont {J.}~\bibnamefont {von
  Neuman}}\ and\ \bibinfo {author} {\bibfnamefont {E.}~\bibnamefont {Wigner}},\
  }\bibfield  {title} {\bibinfo {title} {{\"U}ber merkw{\"u}rdige diskrete
  {Eigenwerte}. {Uber} das {Verhalten} von {Eigenwerten} bei adiabatischen
  {Prozessen}},\ }\href {http://adsabs.harvard.edu/abs/1929PhyZ...30..467V}
  {\bibfield  {journal} {\bibinfo  {journal} {Physikalische Zeitschrift}\
  }\textbf {\bibinfo {volume} {30}},\ \bibinfo {pages} {467} (\bibinfo {year}
  {1929})}\BibitemShut {NoStop}%
\bibitem [{\citenamefont {Friedrich}\ and\ \citenamefont
  {Wintgen}(1985)}]{1985Interfering}%
  \BibitemOpen
  \bibfield  {author} {\bibinfo {author} {\bibfnamefont {H.}~\bibnamefont
  {Friedrich}}\ and\ \bibinfo {author} {\bibfnamefont {D.}~\bibnamefont
  {Wintgen}},\ }\bibfield  {title} {\bibinfo {title} {Interfering resonances
  and bound states in the continuum},\ }\href@noop {} {\bibfield  {journal}
  {\bibinfo  {journal} {Phys. Rev. A}\ }\textbf {\bibinfo {volume} {32}},\
  \bibinfo {pages} {3231} (\bibinfo {year} {1985})}\BibitemShut {NoStop}%
\bibitem [{\citenamefont {Marinica}\ \emph {et~al.}(2008)\citenamefont
  {Marinica}, \citenamefont {Borisov},\ and\ \citenamefont
  {Shabanov}}]{marinica2008bound}%
  \BibitemOpen
  \bibfield  {author} {\bibinfo {author} {\bibfnamefont {D.}~\bibnamefont
  {Marinica}}, \bibinfo {author} {\bibfnamefont {A.}~\bibnamefont {Borisov}},\
  and\ \bibinfo {author} {\bibfnamefont {S.}~\bibnamefont {Shabanov}},\
  }\bibfield  {title} {\bibinfo {title} {Bound states in the continuum in
  photonics},\ }\href@noop {} {\bibfield  {journal} {\bibinfo  {journal} {Phys.
  Rev. Lett.}\ }\textbf {\bibinfo {volume} {100}},\ \bibinfo {pages} {183902}
  (\bibinfo {year} {2008})}\BibitemShut {NoStop}%
\bibitem [{\citenamefont {Hsu}\ \emph {et~al.}(2016)\citenamefont {Hsu},
  \citenamefont {Zhen}, \citenamefont {Stone}, \citenamefont {Joannopoulos},\
  and\ \citenamefont {Solja\v{c}i\'{c}}}]{hsu_bound_2016}%
  \BibitemOpen
  \bibfield  {author} {\bibinfo {author} {\bibfnamefont {C.~W.}\ \bibnamefont
  {Hsu}}, \bibinfo {author} {\bibfnamefont {B.}~\bibnamefont {Zhen}}, \bibinfo
  {author} {\bibfnamefont {A.~D.}\ \bibnamefont {Stone}}, \bibinfo {author}
  {\bibfnamefont {J.~D.}\ \bibnamefont {Joannopoulos}},\ and\ \bibinfo {author}
  {\bibfnamefont {M.}~\bibnamefont {Solja\v{c}i\'{c}}},\ }\bibfield  {title}
  {\bibinfo {title} {Bound states in the continuum},\ }\href
  {https://doi.org/10.1038/natrevmats.2016.48} {\bibfield  {journal} {\bibinfo
  {journal} {Nat. Rev. Mater.}\ }\textbf {\bibinfo {volume} {1}},\ \bibinfo
  {pages} {16048} (\bibinfo {year} {2016})}\BibitemShut {NoStop}%
\bibitem [{\citenamefont {Koshelev}\ \emph {et~al.}(2020)\citenamefont
  {Koshelev}, \citenamefont {Bogdanov}, \citenamefont {Kivshar} \emph
  {et~al.}}]{koshelev2020engineering}%
  \BibitemOpen
  \bibfield  {author} {\bibinfo {author} {\bibfnamefont {K.}~\bibnamefont
  {Koshelev}}, \bibinfo {author} {\bibfnamefont {A.}~\bibnamefont {Bogdanov}},
  \bibinfo {author} {\bibfnamefont {Y.}~\bibnamefont {Kivshar}}, \emph
  {et~al.},\ }\bibfield  {title} {\bibinfo {title} {Engineering with bound
  states in the continuum},\ }\href@noop {} {\bibfield  {journal} {\bibinfo
  {journal} {Opt. Photonics News}\ }\textbf {\bibinfo {volume} {31}},\ \bibinfo
  {pages} {38} (\bibinfo {year} {2020})}\BibitemShut {NoStop}%
\bibitem [{\citenamefont {Sadreev}(2021)}]{sadreev2021interference}%
  \BibitemOpen
  \bibfield  {author} {\bibinfo {author} {\bibfnamefont {A.~F.}\ \bibnamefont
  {Sadreev}},\ }\bibfield  {title} {\bibinfo {title} {Interference traps waves
  in an open system: bound states in the continuum},\ }\href@noop {} {\bibfield
   {journal} {\bibinfo  {journal} {Rep. Prog. Phys.}\ }\textbf {\bibinfo
  {volume} {84}},\ \bibinfo {pages} {055901} (\bibinfo {year}
  {2021})}\BibitemShut {NoStop}%
\bibitem [{\citenamefont {Hu}\ \emph {et~al.}(2022)\citenamefont {Hu},
  \citenamefont {Wang}, \citenamefont {Jiang}, \citenamefont {Wang},
  \citenamefont {Shi}, \citenamefont {Han}, \citenamefont {Zhang},
  \citenamefont {Chan},\ and\ \citenamefont {Zi}}]{hu2022global}%
  \BibitemOpen
  \bibfield  {author} {\bibinfo {author} {\bibfnamefont {P.}~\bibnamefont
  {Hu}}, \bibinfo {author} {\bibfnamefont {J.}~\bibnamefont {Wang}}, \bibinfo
  {author} {\bibfnamefont {Q.}~\bibnamefont {Jiang}}, \bibinfo {author}
  {\bibfnamefont {J.}~\bibnamefont {Wang}}, \bibinfo {author} {\bibfnamefont
  {L.}~\bibnamefont {Shi}}, \bibinfo {author} {\bibfnamefont {D.}~\bibnamefont
  {Han}}, \bibinfo {author} {\bibfnamefont {Z.-Q.}\ \bibnamefont {Zhang}},
  \bibinfo {author} {\bibfnamefont {C.~T.}\ \bibnamefont {Chan}},\ and\
  \bibinfo {author} {\bibfnamefont {J.}~\bibnamefont {Zi}},\ }\bibfield
  {title} {\bibinfo {title} {Global phase diagram of bound states in the
  continuum},\ }\href@noop {} {\bibfield  {journal} {\bibinfo  {journal}
  {Optica}\ }\textbf {\bibinfo {volume} {9}},\ \bibinfo {pages} {1353}
  (\bibinfo {year} {2022})}\BibitemShut {NoStop}%
\bibitem [{\citenamefont {Dennis}\ \emph {et~al.}(2009)\citenamefont {Dennis},
  \citenamefont {O'Holleran},\ and\ \citenamefont
  {Padgett}}]{dennis2009singular}%
  \BibitemOpen
  \bibfield  {author} {\bibinfo {author} {\bibfnamefont {M.~R.}\ \bibnamefont
  {Dennis}}, \bibinfo {author} {\bibfnamefont {K.}~\bibnamefont {O'Holleran}},\
  and\ \bibinfo {author} {\bibfnamefont {M.~J.}\ \bibnamefont {Padgett}},\
  }\bibfield  {title} {\bibinfo {title} {Singular optics: optical vortices and
  polarization singularities},\ }\href@noop {} {\bibfield  {journal} {\bibinfo
  {journal} {Prog. Opt.}\ }\textbf {\bibinfo {volume} {53}},\ \bibinfo {pages}
  {293} (\bibinfo {year} {2009})}\BibitemShut {NoStop}%
\bibitem [{\citenamefont {Soskin}\ \emph {et~al.}(2016)\citenamefont {Soskin},
  \citenamefont {Boriskina}, \citenamefont {Chong}, \citenamefont {Dennis},\
  and\ \citenamefont {Desyatnikov}}]{Soskin_2016}%
  \BibitemOpen
  \bibfield  {author} {\bibinfo {author} {\bibfnamefont {M.}~\bibnamefont
  {Soskin}}, \bibinfo {author} {\bibfnamefont {S.~V.}\ \bibnamefont
  {Boriskina}}, \bibinfo {author} {\bibfnamefont {Y.}~\bibnamefont {Chong}},
  \bibinfo {author} {\bibfnamefont {M.~R.}\ \bibnamefont {Dennis}},\ and\
  \bibinfo {author} {\bibfnamefont {A.}~\bibnamefont {Desyatnikov}},\
  }\bibfield  {title} {\bibinfo {title} {Singular optics and topological
  photonics},\ }\href {https://doi.org/10.1088/2040-8986/19/1/010401}
  {\bibfield  {journal} {\bibinfo  {journal} {J. Opt.}\ }\textbf {\bibinfo
  {volume} {19}},\ \bibinfo {pages} {010401} (\bibinfo {year}
  {2016})}\BibitemShut {NoStop}%
\bibitem [{\citenamefont {Gbur}(2016)}]{gbur_singular_2016}%
  \BibitemOpen
  \bibfield  {author} {\bibinfo {author} {\bibfnamefont {G.~J.}\ \bibnamefont
  {Gbur}},\ }\href@noop {} {\emph {\bibinfo {title} {Singular {Optics}}}}\
  (\bibinfo  {publisher} {CRC Press},\ \bibinfo {year} {2016})\BibitemShut
  {NoStop}%
\bibitem [{\citenamefont {Feng}\ \emph {et~al.}(2017)\citenamefont {Feng},
  \citenamefont {El-Ganainy},\ and\ \citenamefont {Ge}}]{feng2017non}%
  \BibitemOpen
  \bibfield  {author} {\bibinfo {author} {\bibfnamefont {L.}~\bibnamefont
  {Feng}}, \bibinfo {author} {\bibfnamefont {R.}~\bibnamefont {El-Ganainy}},\
  and\ \bibinfo {author} {\bibfnamefont {L.}~\bibnamefont {Ge}},\ }\bibfield
  {title} {\bibinfo {title} {Non-hermitian photonics based on parity--time
  symmetry},\ }\href@noop {} {\bibfield  {journal} {\bibinfo  {journal} {Nat.
  Photonics}\ }\textbf {\bibinfo {volume} {11}},\ \bibinfo {pages} {752}
  (\bibinfo {year} {2017})}\BibitemShut {NoStop}%
\bibitem [{\citenamefont {Leykam}\ \emph {et~al.}(2017)\citenamefont {Leykam},
  \citenamefont {Bliokh}, \citenamefont {Huang}, \citenamefont {Chong},\ and\
  \citenamefont {Nori}}]{leykam2017edge}%
  \BibitemOpen
  \bibfield  {author} {\bibinfo {author} {\bibfnamefont {D.}~\bibnamefont
  {Leykam}}, \bibinfo {author} {\bibfnamefont {K.~Y.}\ \bibnamefont {Bliokh}},
  \bibinfo {author} {\bibfnamefont {C.}~\bibnamefont {Huang}}, \bibinfo
  {author} {\bibfnamefont {Y.~D.}\ \bibnamefont {Chong}},\ and\ \bibinfo
  {author} {\bibfnamefont {F.}~\bibnamefont {Nori}},\ }\bibfield  {title}
  {\bibinfo {title} {Edge modes, degeneracies, and topological numbers in
  non-hermitian systems},\ }\href@noop {} {\bibfield  {journal} {\bibinfo
  {journal} {Phys. Rev. Lett.}\ }\textbf {\bibinfo {volume} {118}},\ \bibinfo
  {pages} {040401} (\bibinfo {year} {2017})}\BibitemShut {NoStop}%
\bibitem [{\citenamefont {El-Ganainy}\ \emph {et~al.}(2018)\citenamefont
  {El-Ganainy}, \citenamefont {Makris}, \citenamefont {Khajavikhan},
  \citenamefont {Musslimani}, \citenamefont {Rotter},\ and\ \citenamefont
  {Christodoulides}}]{el2018non}%
  \BibitemOpen
  \bibfield  {author} {\bibinfo {author} {\bibfnamefont {R.}~\bibnamefont
  {El-Ganainy}}, \bibinfo {author} {\bibfnamefont {K.~G.}\ \bibnamefont
  {Makris}}, \bibinfo {author} {\bibfnamefont {M.}~\bibnamefont {Khajavikhan}},
  \bibinfo {author} {\bibfnamefont {Z.~H.}\ \bibnamefont {Musslimani}},
  \bibinfo {author} {\bibfnamefont {S.}~\bibnamefont {Rotter}},\ and\ \bibinfo
  {author} {\bibfnamefont {D.~N.}\ \bibnamefont {Christodoulides}},\ }\bibfield
   {title} {\bibinfo {title} {Non-hermitian physics and pt symmetry},\
  }\href@noop {} {\bibfield  {journal} {\bibinfo  {journal} {Nat. Phys.}\
  }\textbf {\bibinfo {volume} {14}},\ \bibinfo {pages} {11} (\bibinfo {year}
  {2018})}\BibitemShut {NoStop}%
\bibitem [{\citenamefont {Shen}\ \emph {et~al.}(2018)\citenamefont {Shen},
  \citenamefont {Zhen},\ and\ \citenamefont {Fu}}]{shen2018topological}%
  \BibitemOpen
  \bibfield  {author} {\bibinfo {author} {\bibfnamefont {H.}~\bibnamefont
  {Shen}}, \bibinfo {author} {\bibfnamefont {B.}~\bibnamefont {Zhen}},\ and\
  \bibinfo {author} {\bibfnamefont {L.}~\bibnamefont {Fu}},\ }\bibfield
  {title} {\bibinfo {title} {Topological band theory for non-hermitian
  hamiltonians},\ }\href@noop {} {\bibfield  {journal} {\bibinfo  {journal}
  {Phys. Rev. Lett.}\ }\textbf {\bibinfo {volume} {120}},\ \bibinfo {pages}
  {146402} (\bibinfo {year} {2018})}\BibitemShut {NoStop}%
\bibitem [{\citenamefont {Bergholtz}\ \emph {et~al.}(2021)\citenamefont
  {Bergholtz}, \citenamefont {Budich},\ and\ \citenamefont
  {Kunst}}]{bergholtz2021exceptional}%
  \BibitemOpen
  \bibfield  {author} {\bibinfo {author} {\bibfnamefont {E.~J.}\ \bibnamefont
  {Bergholtz}}, \bibinfo {author} {\bibfnamefont {J.~C.}\ \bibnamefont
  {Budich}},\ and\ \bibinfo {author} {\bibfnamefont {F.~K.}\ \bibnamefont
  {Kunst}},\ }\bibfield  {title} {\bibinfo {title} {Exceptional topology of
  non-hermitian systems},\ }\href@noop {} {\bibfield  {journal} {\bibinfo
  {journal} {Rev. Mod. Phys.}\ }\textbf {\bibinfo {volume} {93}},\ \bibinfo
  {pages} {015005} (\bibinfo {year} {2021})}\BibitemShut {NoStop}%
\bibitem [{\citenamefont {Chen}\ \emph {et~al.}(2022)\citenamefont {Chen},
  \citenamefont {Yin}, \citenamefont {Jin}, \citenamefont {Zheng},
  \citenamefont {Zhang}, \citenamefont {Wang}, \citenamefont {He},
  \citenamefont {Zhen},\ and\ \citenamefont {Peng}}]{chen2022observation}%
  \BibitemOpen
  \bibfield  {author} {\bibinfo {author} {\bibfnamefont {Z.}~\bibnamefont
  {Chen}}, \bibinfo {author} {\bibfnamefont {X.}~\bibnamefont {Yin}}, \bibinfo
  {author} {\bibfnamefont {J.}~\bibnamefont {Jin}}, \bibinfo {author}
  {\bibfnamefont {Z.}~\bibnamefont {Zheng}}, \bibinfo {author} {\bibfnamefont
  {Z.}~\bibnamefont {Zhang}}, \bibinfo {author} {\bibfnamefont
  {F.}~\bibnamefont {Wang}}, \bibinfo {author} {\bibfnamefont {L.}~\bibnamefont
  {He}}, \bibinfo {author} {\bibfnamefont {B.}~\bibnamefont {Zhen}},\ and\
  \bibinfo {author} {\bibfnamefont {C.}~\bibnamefont {Peng}},\ }\bibfield
  {title} {\bibinfo {title} {Observation of miniaturized bound states in the
  continuum with ultra-high quality factors},\ }\href@noop {} {\bibfield
  {journal} {\bibinfo  {journal} {Sci. Bull.}\ }\textbf {\bibinfo {volume}
  {67}},\ \bibinfo {pages} {359} (\bibinfo {year} {2022})}\BibitemShut
  {NoStop}%
\bibitem [{\citenamefont {Huang}\ \emph {et~al.}(2020)\citenamefont {Huang},
  \citenamefont {Zhang}, \citenamefont {Xiao}, \citenamefont {Wang},
  \citenamefont {Fan}, \citenamefont {Liu}, \citenamefont {Zhang},
  \citenamefont {Qu}, \citenamefont {Ji}, \citenamefont {Han} \emph
  {et~al.}}]{huang2020ultrafast}%
  \BibitemOpen
  \bibfield  {author} {\bibinfo {author} {\bibfnamefont {C.}~\bibnamefont
  {Huang}}, \bibinfo {author} {\bibfnamefont {C.}~\bibnamefont {Zhang}},
  \bibinfo {author} {\bibfnamefont {S.}~\bibnamefont {Xiao}}, \bibinfo {author}
  {\bibfnamefont {Y.}~\bibnamefont {Wang}}, \bibinfo {author} {\bibfnamefont
  {Y.}~\bibnamefont {Fan}}, \bibinfo {author} {\bibfnamefont {Y.}~\bibnamefont
  {Liu}}, \bibinfo {author} {\bibfnamefont {N.}~\bibnamefont {Zhang}}, \bibinfo
  {author} {\bibfnamefont {G.}~\bibnamefont {Qu}}, \bibinfo {author}
  {\bibfnamefont {H.}~\bibnamefont {Ji}}, \bibinfo {author} {\bibfnamefont
  {J.}~\bibnamefont {Han}}, \emph {et~al.},\ }\bibfield  {title} {\bibinfo
  {title} {Ultrafast control of vortex microlasers},\ }\href@noop {} {\bibfield
   {journal} {\bibinfo  {journal} {Science}\ }\textbf {\bibinfo {volume}
  {367}},\ \bibinfo {pages} {1018} (\bibinfo {year} {2020})}\BibitemShut
  {NoStop}%
\bibitem [{\citenamefont {Zhang}\ \emph {et~al.}(2022)\citenamefont {Zhang},
  \citenamefont {Liu}, \citenamefont {Han}, \citenamefont {Kivshar},\ and\
  \citenamefont {Song}}]{zhang2022chiral}%
  \BibitemOpen
  \bibfield  {author} {\bibinfo {author} {\bibfnamefont {X.}~\bibnamefont
  {Zhang}}, \bibinfo {author} {\bibfnamefont {Y.}~\bibnamefont {Liu}}, \bibinfo
  {author} {\bibfnamefont {J.}~\bibnamefont {Han}}, \bibinfo {author}
  {\bibfnamefont {Y.}~\bibnamefont {Kivshar}},\ and\ \bibinfo {author}
  {\bibfnamefont {Q.}~\bibnamefont {Song}},\ }\bibfield  {title} {\bibinfo
  {title} {Chiral emission from resonant metasurfaces},\ }\href@noop {}
  {\bibfield  {journal} {\bibinfo  {journal} {Science}\ }\textbf {\bibinfo
  {volume} {377}},\ \bibinfo {pages} {1215} (\bibinfo {year}
  {2022})}\BibitemShut {NoStop}%
\bibitem [{\citenamefont {Chen}\ \emph {et~al.}(2023)\citenamefont {Chen},
  \citenamefont {Deng}, \citenamefont {Sha}, \citenamefont {Chen},
  \citenamefont {Wang}, \citenamefont {Chen}, \citenamefont {Wu}, \citenamefont
  {Chu}, \citenamefont {Kivshar}, \citenamefont {Xiao} \emph
  {et~al.}}]{chen2023observation}%
  \BibitemOpen
  \bibfield  {author} {\bibinfo {author} {\bibfnamefont {Y.}~\bibnamefont
  {Chen}}, \bibinfo {author} {\bibfnamefont {H.}~\bibnamefont {Deng}}, \bibinfo
  {author} {\bibfnamefont {X.}~\bibnamefont {Sha}}, \bibinfo {author}
  {\bibfnamefont {W.}~\bibnamefont {Chen}}, \bibinfo {author} {\bibfnamefont
  {R.}~\bibnamefont {Wang}}, \bibinfo {author} {\bibfnamefont {Y.-H.}\
  \bibnamefont {Chen}}, \bibinfo {author} {\bibfnamefont {D.}~\bibnamefont
  {Wu}}, \bibinfo {author} {\bibfnamefont {J.}~\bibnamefont {Chu}}, \bibinfo
  {author} {\bibfnamefont {Y.~S.}\ \bibnamefont {Kivshar}}, \bibinfo {author}
  {\bibfnamefont {S.}~\bibnamefont {Xiao}}, \emph {et~al.},\ }\bibfield
  {title} {\bibinfo {title} {Observation of intrinsic chiral bound states in
  the continuum},\ }\href@noop {} {\bibfield  {journal} {\bibinfo  {journal}
  {Nature}\ }\textbf {\bibinfo {volume} {613}},\ \bibinfo {pages} {474}
  (\bibinfo {year} {2023})}\BibitemShut {NoStop}%
\bibitem [{\citenamefont {Lu}\ \emph {et~al.}(2014)\citenamefont {Lu},
  \citenamefont {Joannopoulos},\ and\ \citenamefont
  {Soljačić}}]{lu_topological_2014}%
  \BibitemOpen
  \bibfield  {author} {\bibinfo {author} {\bibfnamefont {L.}~\bibnamefont
  {Lu}}, \bibinfo {author} {\bibfnamefont {J.~D.}\ \bibnamefont
  {Joannopoulos}},\ and\ \bibinfo {author} {\bibfnamefont {M.}~\bibnamefont
  {Soljačić}},\ }\bibfield  {title} {\bibinfo {title} {Topological
  photonics},\ }\href {https://doi.org/10.1038/nphoton.2014.248} {\bibfield
  {journal} {\bibinfo  {journal} {Nature Photon.}\ }\textbf {\bibinfo {volume}
  {8}},\ \bibinfo {pages} {821} (\bibinfo {year} {2014})}\BibitemShut {NoStop}%
\bibitem [{\citenamefont {Khanikaev}\ and\ \citenamefont
  {Shvets}(2017)}]{khanikaev_two-dimensional_2017}%
  \BibitemOpen
  \bibfield  {author} {\bibinfo {author} {\bibfnamefont {A.~B.}\ \bibnamefont
  {Khanikaev}}\ and\ \bibinfo {author} {\bibfnamefont {G.}~\bibnamefont
  {Shvets}},\ }\bibfield  {title} {\bibinfo {title} {Two-dimensional
  topological photonics},\ }\href {https://doi.org/10.1038/s41566-017-0048-5}
  {\bibfield  {journal} {\bibinfo  {journal} {Nature Photon.}\ }\textbf
  {\bibinfo {volume} {11}},\ \bibinfo {pages} {763} (\bibinfo {year}
  {2017})}\BibitemShut {NoStop}%
\bibitem [{\citenamefont {Ozawa}\ \emph {et~al.}(2019)\citenamefont {Ozawa},
  \citenamefont {Price}, \citenamefont {Amo}, \citenamefont {Goldman},
  \citenamefont {Hafezi}, \citenamefont {Lu}, \citenamefont {Rechtsman},
  \citenamefont {Schuster}, \citenamefont {Simon}, \citenamefont {Zilberberg}
  \emph {et~al.}}]{ozawa_topological_2019}%
  \BibitemOpen
  \bibfield  {author} {\bibinfo {author} {\bibfnamefont {T.}~\bibnamefont
  {Ozawa}}, \bibinfo {author} {\bibfnamefont {H.~M.}\ \bibnamefont {Price}},
  \bibinfo {author} {\bibfnamefont {A.}~\bibnamefont {Amo}}, \bibinfo {author}
  {\bibfnamefont {N.}~\bibnamefont {Goldman}}, \bibinfo {author} {\bibfnamefont
  {M.}~\bibnamefont {Hafezi}}, \bibinfo {author} {\bibfnamefont
  {L.}~\bibnamefont {Lu}}, \bibinfo {author} {\bibfnamefont {M.~C.}\
  \bibnamefont {Rechtsman}}, \bibinfo {author} {\bibfnamefont {D.}~\bibnamefont
  {Schuster}}, \bibinfo {author} {\bibfnamefont {J.}~\bibnamefont {Simon}},
  \bibinfo {author} {\bibfnamefont {O.}~\bibnamefont {Zilberberg}}, \emph
  {et~al.},\ }\bibfield  {title} {\bibinfo {title} {Topological photonics},\
  }\href {https://doi.org/10.1103/RevModPhys.91.015006} {\bibfield  {journal}
  {\bibinfo  {journal} {Rev. Mod. Phys.}\ }\textbf {\bibinfo {volume} {91}},\
  \bibinfo {pages} {015006} (\bibinfo {year} {2019})}\BibitemShut {NoStop}%
\bibitem [{\citenamefont {Wang}\ \emph {et~al.}(2020)\citenamefont {Wang},
  \citenamefont {Gupta}, \citenamefont {Xie},\ and\ \citenamefont
  {Lu}}]{wang_topological_2020}%
  \BibitemOpen
  \bibfield  {author} {\bibinfo {author} {\bibfnamefont {H.}~\bibnamefont
  {Wang}}, \bibinfo {author} {\bibfnamefont {S.~K.}\ \bibnamefont {Gupta}},
  \bibinfo {author} {\bibfnamefont {B.}~\bibnamefont {Xie}},\ and\ \bibinfo
  {author} {\bibfnamefont {M.}~\bibnamefont {Lu}},\ }\bibfield  {title}
  {\bibinfo {title} {Topological photonic crystals: a review},\ }\href
  {https://doi.org/10.1007/s12200-019-0949-7} {\bibfield  {journal} {\bibinfo
  {journal} {Front. Optoelectron.}\ }\textbf {\bibinfo {volume} {13}},\
  \bibinfo {pages} {50} (\bibinfo {year} {2020})}\BibitemShut {NoStop}%
\bibitem [{\citenamefont {Zhen}\ \emph {et~al.}(2014)\citenamefont {Zhen},
  \citenamefont {Hsu}, \citenamefont {Lu}, \citenamefont {Stone},\ and\
  \citenamefont {Solja\v{c}i\'{c}}}]{zhen_topological_2014}%
  \BibitemOpen
  \bibfield  {author} {\bibinfo {author} {\bibfnamefont {B.}~\bibnamefont
  {Zhen}}, \bibinfo {author} {\bibfnamefont {C.~W.}\ \bibnamefont {Hsu}},
  \bibinfo {author} {\bibfnamefont {L.}~\bibnamefont {Lu}}, \bibinfo {author}
  {\bibfnamefont {A.~D.}\ \bibnamefont {Stone}},\ and\ \bibinfo {author}
  {\bibfnamefont {M.}~\bibnamefont {Solja\v{c}i\'{c}}},\ }\bibfield  {title}
  {\bibinfo {title} {Topological {nature} of {optical} {bound} {states} in the
  {continuum}},\ }\href {https://doi.org/10.1103/PhysRevLett.113.257401}
  {\bibfield  {journal} {\bibinfo  {journal} {Phys. Rev. Lett.}\ }\textbf
  {\bibinfo {volume} {113}},\ \bibinfo {pages} {257401} (\bibinfo {year}
  {2014})}\BibitemShut {NoStop}%
\bibitem [{\citenamefont {Bulgakov}\ and\ \citenamefont
  {Maksimov}(2017)}]{bulgakov2017bound}%
  \BibitemOpen
  \bibfield  {author} {\bibinfo {author} {\bibfnamefont {E.~N.}\ \bibnamefont
  {Bulgakov}}\ and\ \bibinfo {author} {\bibfnamefont {D.~N.}\ \bibnamefont
  {Maksimov}},\ }\bibfield  {title} {\bibinfo {title} {Bound states in the
  continuum and polarization singularities in periodic arrays of dielectric
  rods},\ }\href@noop {} {\bibfield  {journal} {\bibinfo  {journal} {Phys. Rev.
  A}\ }\textbf {\bibinfo {volume} {96}},\ \bibinfo {pages} {063833} (\bibinfo
  {year} {2017})}\BibitemShut {NoStop}%
\bibitem [{\citenamefont {Zhang}\ \emph {et~al.}(2018)\citenamefont {Zhang},
  \citenamefont {Chen}, \citenamefont {Liu}, \citenamefont {Hsu}, \citenamefont
  {Wang}, \citenamefont {Guan}, \citenamefont {Liu}, \citenamefont {Shi},
  \citenamefont {Lu},\ and\ \citenamefont {Zi}}]{shilei_observation_2018}%
  \BibitemOpen
  \bibfield  {author} {\bibinfo {author} {\bibfnamefont {Y.}~\bibnamefont
  {Zhang}}, \bibinfo {author} {\bibfnamefont {A.}~\bibnamefont {Chen}},
  \bibinfo {author} {\bibfnamefont {W.}~\bibnamefont {Liu}}, \bibinfo {author}
  {\bibfnamefont {C.~W.}\ \bibnamefont {Hsu}}, \bibinfo {author} {\bibfnamefont
  {B.}~\bibnamefont {Wang}}, \bibinfo {author} {\bibfnamefont {F.}~\bibnamefont
  {Guan}}, \bibinfo {author} {\bibfnamefont {X.}~\bibnamefont {Liu}}, \bibinfo
  {author} {\bibfnamefont {L.}~\bibnamefont {Shi}}, \bibinfo {author}
  {\bibfnamefont {L.}~\bibnamefont {Lu}},\ and\ \bibinfo {author}
  {\bibfnamefont {J.}~\bibnamefont {Zi}},\ }\bibfield  {title} {\bibinfo
  {title} {Observation of polarization vortices in momentum space},\ }\href
  {https://doi.org/10.1103/PhysRevLett.120.186103} {\bibfield  {journal}
  {\bibinfo  {journal} {Phys. Rev. Lett.}\ }\textbf {\bibinfo {volume} {120}},\
  \bibinfo {pages} {186103} (\bibinfo {year} {2018})}\BibitemShut {NoStop}%
\bibitem [{\citenamefont {Doeleman}\ \emph {et~al.}(2018)\citenamefont
  {Doeleman}, \citenamefont {Monticone}, \citenamefont {den Hollander},
  \citenamefont {Alu},\ and\ \citenamefont
  {Koenderink}}]{alu_experimental_2018}%
  \BibitemOpen
  \bibfield  {author} {\bibinfo {author} {\bibfnamefont {H.~M.}\ \bibnamefont
  {Doeleman}}, \bibinfo {author} {\bibfnamefont {F.}~\bibnamefont {Monticone}},
  \bibinfo {author} {\bibfnamefont {W.}~\bibnamefont {den Hollander}}, \bibinfo
  {author} {\bibfnamefont {A.}~\bibnamefont {Alu}},\ and\ \bibinfo {author}
  {\bibfnamefont {A.~F.}\ \bibnamefont {Koenderink}},\ }\bibfield  {title}
  {\bibinfo {title} {Experimental observation of a polarization vortex at an
  optical bound state in the continuum},\ }\href
  {https://doi.org/10.1038/s41566-018-0177-5} {\bibfield  {journal} {\bibinfo
  {journal} {Nat. Photonics}\ }\textbf {\bibinfo {volume} {12}},\ \bibinfo
  {pages} {397} (\bibinfo {year} {2018})}\BibitemShut {NoStop}%
\bibitem [{\citenamefont {Chen}\ \emph {et~al.}(2021)\citenamefont {Chen},
  \citenamefont {Yang}, \citenamefont {Chen},\ and\ \citenamefont
  {Liu}}]{liuwei_global_charge}%
  \BibitemOpen
  \bibfield  {author} {\bibinfo {author} {\bibfnamefont {W.}~\bibnamefont
  {Chen}}, \bibinfo {author} {\bibfnamefont {Q.}~\bibnamefont {Yang}}, \bibinfo
  {author} {\bibfnamefont {Y.}~\bibnamefont {Chen}},\ and\ \bibinfo {author}
  {\bibfnamefont {W.}~\bibnamefont {Liu}},\ }\bibfield  {title} {\bibinfo
  {title} {Evolution and global charge conservation for polarization
  singularities emerging from non-hermitian degeneracies},\ }\href
  {https://doi.org/10.1073/pnas.2019578118} {\bibfield  {journal} {\bibinfo
  {journal} {Proc. Natl. Acad. Sci. U.S.A.}\ }\textbf {\bibinfo {volume}
  {118}},\ \bibinfo {pages} {e2019578118} (\bibinfo {year} {2021})}\BibitemShut
  {NoStop}%
\bibitem [{\citenamefont {Che}\ \emph {et~al.}(2021)\citenamefont {Che},
  \citenamefont {Zhang}, \citenamefont {Liu}, \citenamefont {Zhao},
  \citenamefont {Wang}, \citenamefont {Zhang}, \citenamefont {Guan},
  \citenamefont {Liu}, \citenamefont {Liu}, \citenamefont {Shi} \emph
  {et~al.}}]{shilei_polarization_singularity_2021}%
  \BibitemOpen
  \bibfield  {author} {\bibinfo {author} {\bibfnamefont {Z.}~\bibnamefont
  {Che}}, \bibinfo {author} {\bibfnamefont {Y.}~\bibnamefont {Zhang}}, \bibinfo
  {author} {\bibfnamefont {W.}~\bibnamefont {Liu}}, \bibinfo {author}
  {\bibfnamefont {M.}~\bibnamefont {Zhao}}, \bibinfo {author} {\bibfnamefont
  {J.}~\bibnamefont {Wang}}, \bibinfo {author} {\bibfnamefont {W.}~\bibnamefont
  {Zhang}}, \bibinfo {author} {\bibfnamefont {F.}~\bibnamefont {Guan}},
  \bibinfo {author} {\bibfnamefont {X.}~\bibnamefont {Liu}}, \bibinfo {author}
  {\bibfnamefont {W.}~\bibnamefont {Liu}}, \bibinfo {author} {\bibfnamefont
  {L.}~\bibnamefont {Shi}}, \emph {et~al.},\ }\bibfield  {title} {\bibinfo
  {title} {Polarization singularities of photonic quasicrystals in momentum
  space},\ }\href {https://doi.org/10.1103/PhysRevLett.127.043901} {\bibfield
  {journal} {\bibinfo  {journal} {Phys. Rev. Lett.}\ }\textbf {\bibinfo
  {volume} {127}},\ \bibinfo {pages} {043901} (\bibinfo {year}
  {2021})}\BibitemShut {NoStop}%
\bibitem [{\citenamefont {Jin}\ \emph {et~al.}(2019)\citenamefont {Jin},
  \citenamefont {Yin}, \citenamefont {Ni}, \citenamefont {Soljačić},
  \citenamefont {Zhen},\ and\ \citenamefont {Peng}}]{jin_topologically_2019}%
  \BibitemOpen
  \bibfield  {author} {\bibinfo {author} {\bibfnamefont {J.}~\bibnamefont
  {Jin}}, \bibinfo {author} {\bibfnamefont {X.}~\bibnamefont {Yin}}, \bibinfo
  {author} {\bibfnamefont {L.}~\bibnamefont {Ni}}, \bibinfo {author}
  {\bibfnamefont {M.}~\bibnamefont {Soljačić}}, \bibinfo {author}
  {\bibfnamefont {B.}~\bibnamefont {Zhen}},\ and\ \bibinfo {author}
  {\bibfnamefont {C.}~\bibnamefont {Peng}},\ }\bibfield  {title} {\bibinfo
  {title} {Topologically enabled ultrahigh-q guided resonances robust to
  out-of-plane scattering},\ }\href {https://doi.org/10.1038/s41586-019-1664-7}
  {\bibfield  {journal} {\bibinfo  {journal} {Nature}\ }\textbf {\bibinfo
  {volume} {574}},\ \bibinfo {pages} {501} (\bibinfo {year}
  {2019})}\BibitemShut {NoStop}%
\bibitem [{\citenamefont {Kang}\ \emph {et~al.}(2021)\citenamefont {Kang},
  \citenamefont {Zhang}, \citenamefont {Xiao},\ and\ \citenamefont
  {Xu}}]{kang2021merging}%
  \BibitemOpen
  \bibfield  {author} {\bibinfo {author} {\bibfnamefont {M.}~\bibnamefont
  {Kang}}, \bibinfo {author} {\bibfnamefont {S.}~\bibnamefont {Zhang}},
  \bibinfo {author} {\bibfnamefont {M.}~\bibnamefont {Xiao}},\ and\ \bibinfo
  {author} {\bibfnamefont {H.}~\bibnamefont {Xu}},\ }\bibfield  {title}
  {\bibinfo {title} {Merging bound states in the continuum at off-high symmetry
  points},\ }\href@noop {} {\bibfield  {journal} {\bibinfo  {journal} {Phys.
  Rev. Lett.}\ }\textbf {\bibinfo {volume} {126}},\ \bibinfo {pages} {117402}
  (\bibinfo {year} {2021})}\BibitemShut {NoStop}%
\bibitem [{\citenamefont {Hwang}\ \emph {et~al.}(2021)\citenamefont {Hwang},
  \citenamefont {Lee}, \citenamefont {Kim}, \citenamefont {Jeong},
  \citenamefont {Kwon}, \citenamefont {Koshelev}, \citenamefont {Kivshar},\
  and\ \citenamefont {Park}}]{hwang2021ultralow}%
  \BibitemOpen
  \bibfield  {author} {\bibinfo {author} {\bibfnamefont {M.-S.}\ \bibnamefont
  {Hwang}}, \bibinfo {author} {\bibfnamefont {H.-C.}\ \bibnamefont {Lee}},
  \bibinfo {author} {\bibfnamefont {K.-H.}\ \bibnamefont {Kim}}, \bibinfo
  {author} {\bibfnamefont {K.-Y.}\ \bibnamefont {Jeong}}, \bibinfo {author}
  {\bibfnamefont {S.-H.}\ \bibnamefont {Kwon}}, \bibinfo {author}
  {\bibfnamefont {K.}~\bibnamefont {Koshelev}}, \bibinfo {author}
  {\bibfnamefont {Y.}~\bibnamefont {Kivshar}},\ and\ \bibinfo {author}
  {\bibfnamefont {H.-G.}\ \bibnamefont {Park}},\ }\bibfield  {title} {\bibinfo
  {title} {Ultralow-threshold laser using super-bound states in the
  continuum},\ }\href@noop {} {\bibfield  {journal} {\bibinfo  {journal} {Nat.
  Commun.}\ }\textbf {\bibinfo {volume} {12}},\ \bibinfo {pages} {4135}
  (\bibinfo {year} {2021})}\BibitemShut {NoStop}%
\bibitem [{\citenamefont {Yin}\ \emph {et~al.}(2020)\citenamefont {Yin},
  \citenamefont {Jin}, \citenamefont {Soljačić}, \citenamefont {Peng},\ and\
  \citenamefont {Zhen}}]{yin_observation_2020}%
  \BibitemOpen
  \bibfield  {author} {\bibinfo {author} {\bibfnamefont {X.}~\bibnamefont
  {Yin}}, \bibinfo {author} {\bibfnamefont {J.}~\bibnamefont {Jin}}, \bibinfo
  {author} {\bibfnamefont {M.}~\bibnamefont {Soljačić}}, \bibinfo {author}
  {\bibfnamefont {C.}~\bibnamefont {Peng}},\ and\ \bibinfo {author}
  {\bibfnamefont {B.}~\bibnamefont {Zhen}},\ }\bibfield  {title} {\bibinfo
  {title} {Observation of topologically enabled unidirectional guided
  resonances},\ }\href {https://doi.org/10.1038/s41586-020-2181-4} {\bibfield
  {journal} {\bibinfo  {journal} {Nature}\ }\textbf {\bibinfo {volume} {580}},\
  \bibinfo {pages} {467} (\bibinfo {year} {2020})}\BibitemShut {NoStop}%
\bibitem [{\citenamefont {Zeng}\ \emph {et~al.}(2021)\citenamefont {Zeng},
  \citenamefont {Hu}, \citenamefont {Liu}, \citenamefont {Tang},\ and\
  \citenamefont {Qiu}}]{zeng_dynamic_2021}%
  \BibitemOpen
  \bibfield  {author} {\bibinfo {author} {\bibfnamefont {Y.}~\bibnamefont
  {Zeng}}, \bibinfo {author} {\bibfnamefont {G.}~\bibnamefont {Hu}}, \bibinfo
  {author} {\bibfnamefont {K.}~\bibnamefont {Liu}}, \bibinfo {author}
  {\bibfnamefont {Z.}~\bibnamefont {Tang}},\ and\ \bibinfo {author}
  {\bibfnamefont {C.-W.}\ \bibnamefont {Qiu}},\ }\bibfield  {title} {\bibinfo
  {title} {Dynamics of topological polarization singularity in momentum
  space},\ }\href {https://doi.org/10.1103/PhysRevLett.127.176101} {\bibfield
  {journal} {\bibinfo  {journal} {Phys. Rev. Lett.}\ }\textbf {\bibinfo
  {volume} {127}},\ \bibinfo {pages} {176101} (\bibinfo {year}
  {2021})}\BibitemShut {NoStop}%
\bibitem [{\citenamefont {Yin}\ \emph {et~al.}(2023)\citenamefont {Yin},
  \citenamefont {Inoue}, \citenamefont {Peng},\ and\ \citenamefont
  {Noda}}]{yin2023topological}%
  \BibitemOpen
  \bibfield  {author} {\bibinfo {author} {\bibfnamefont {X.}~\bibnamefont
  {Yin}}, \bibinfo {author} {\bibfnamefont {T.}~\bibnamefont {Inoue}}, \bibinfo
  {author} {\bibfnamefont {C.}~\bibnamefont {Peng}},\ and\ \bibinfo {author}
  {\bibfnamefont {S.}~\bibnamefont {Noda}},\ }\bibfield  {title} {\bibinfo
  {title} {Topological unidirectional guided resonances emerged from interband
  coupling},\ }\href@noop {} {\bibfield  {journal} {\bibinfo  {journal} {Phys.
  Rev. Lett.}\ }\textbf {\bibinfo {volume} {130}},\ \bibinfo {pages} {056401}
  (\bibinfo {year} {2023})}\BibitemShut {NoStop}%
\bibitem [{sup()}]{supp}%
  \BibitemOpen
  \bibfield  {title} {\bibinfo {title} {See supplemental material for
  analytical derivation upon lattice coupling and formation of lattice charge,
  detailed evolution of polarization defects induced by inter-band coupling,
  examples and discussions for polarization defects in 2d phc slab and super
  lattice, which includes refs. [51,54,60,6]},\ }\href@noop {} {\ }\BibitemShut
  {NoStop}%
\bibitem [{\citenamefont {Kogelnik}\ and\ \citenamefont
  {Shank}(1972)}]{kogelnik1972coupled}%
  \BibitemOpen
  \bibfield  {author} {\bibinfo {author} {\bibfnamefont {H.}~\bibnamefont
  {Kogelnik}}\ and\ \bibinfo {author} {\bibfnamefont {C.~V.}\ \bibnamefont
  {Shank}},\ }\bibfield  {title} {\bibinfo {title} {Coupled-wave theory of
  distributed feedback lasers},\ }\href@noop {} {\bibfield  {journal} {\bibinfo
   {journal} {J. Appl. Phys.}\ }\textbf {\bibinfo {volume} {43}},\ \bibinfo
  {pages} {2327} (\bibinfo {year} {1972})}\BibitemShut {NoStop}%
\bibitem [{\citenamefont {Liang}\ \emph {et~al.}(2011)\citenamefont {Liang},
  \citenamefont {Peng}, \citenamefont {Sakai}, \citenamefont {Iwahashi},\ and\
  \citenamefont {Noda}}]{liang_three-dimensional_2011}%
  \BibitemOpen
  \bibfield  {author} {\bibinfo {author} {\bibfnamefont {Y.}~\bibnamefont
  {Liang}}, \bibinfo {author} {\bibfnamefont {C.}~\bibnamefont {Peng}},
  \bibinfo {author} {\bibfnamefont {K.}~\bibnamefont {Sakai}}, \bibinfo
  {author} {\bibfnamefont {S.}~\bibnamefont {Iwahashi}},\ and\ \bibinfo
  {author} {\bibfnamefont {S.}~\bibnamefont {Noda}},\ }\bibfield  {title}
  {\bibinfo {title} {Three-dimensional coupled-wave model for square-lattice
  photonic crystal lasers with transverse electric polarization: {A} general
  approach},\ }\href {https://doi.org/10.1103/PhysRevB.84.195119} {\bibfield
  {journal} {\bibinfo  {journal} {Phys. Rev. B}\ }\textbf {\bibinfo {volume}
  {84}},\ \bibinfo {pages} {195119} (\bibinfo {year} {2011})}\BibitemShut
  {NoStop}%
\bibitem [{\citenamefont {Abud}\ and\ \citenamefont
  {Sartori}(1983)}]{abud1983geometry}%
  \BibitemOpen
  \bibfield  {author} {\bibinfo {author} {\bibfnamefont {M.}~\bibnamefont
  {Abud}}\ and\ \bibinfo {author} {\bibfnamefont {G.}~\bibnamefont {Sartori}},\
  }\bibfield  {title} {\bibinfo {title} {The geometry of spontaneous symmetry
  breaking},\ }\href@noop {} {\bibfield  {journal} {\bibinfo  {journal} {Ann
  Phys (N Y)}\ }\textbf {\bibinfo {volume} {150}},\ \bibinfo {pages} {307}
  (\bibinfo {year} {1983})}\BibitemShut {NoStop}%
\bibitem [{\citenamefont {Beekman}\ \emph {et~al.}(2019)\citenamefont
  {Beekman}, \citenamefont {Rademaker},\ and\ \citenamefont {van
  Wezel}}]{beekman2019introduction}%
  \BibitemOpen
  \bibfield  {author} {\bibinfo {author} {\bibfnamefont {A.}~\bibnamefont
  {Beekman}}, \bibinfo {author} {\bibfnamefont {L.}~\bibnamefont {Rademaker}},\
  and\ \bibinfo {author} {\bibfnamefont {J.}~\bibnamefont {van Wezel}},\
  }\bibfield  {title} {\bibinfo {title} {An introduction to spontaneous
  symmetry breaking},\ }\href@noop {} {\bibfield  {journal} {\bibinfo
  {journal} {SciPost Phys. Lect. Notes}\ ,\ \bibinfo {pages} {011}} (\bibinfo
  {year} {2019})}\BibitemShut {NoStop}%
\bibitem [{\citenamefont {Kato}(2013)}]{kato2013perturbation}%
  \BibitemOpen
  \bibfield  {author} {\bibinfo {author} {\bibfnamefont {T.}~\bibnamefont
  {Kato}},\ }\href@noop {} {\emph {\bibinfo {title} {Perturbation theory for
  linear operators}}},\ Vol.\ \bibinfo {volume} {132}\ (\bibinfo  {publisher}
  {Springer Science \& Business Media},\ \bibinfo {year} {2013})\BibitemShut
  {NoStop}%
\bibitem [{\citenamefont {Berry}(2004)}]{berry2004physics}%
  \BibitemOpen
  \bibfield  {author} {\bibinfo {author} {\bibfnamefont {M.~V.}\ \bibnamefont
  {Berry}},\ }\bibfield  {title} {\bibinfo {title} {Physics of nonhermitian
  degeneracies},\ }\href@noop {} {\bibfield  {journal} {\bibinfo  {journal}
  {Czechoslov. J. Phys.}\ }\textbf {\bibinfo {volume} {54}},\ \bibinfo {pages}
  {1039} (\bibinfo {year} {2004})}\BibitemShut {NoStop}%
\bibitem [{\citenamefont {Heiss}(2012)}]{heiss2012physics}%
  \BibitemOpen
  \bibfield  {author} {\bibinfo {author} {\bibfnamefont {W.}~\bibnamefont
  {Heiss}},\ }\bibfield  {title} {\bibinfo {title} {The physics of exceptional
  points},\ }\href@noop {} {\bibfield  {journal} {\bibinfo  {journal} {J. Phys.
  Math. Gen.}\ }\textbf {\bibinfo {volume} {45}},\ \bibinfo {pages} {444016}
  (\bibinfo {year} {2012})}\BibitemShut {NoStop}%
\bibitem [{\citenamefont {Zhen}\ \emph {et~al.}(2015)\citenamefont {Zhen},
  \citenamefont {Hsu}, \citenamefont {Igarashi}, \citenamefont {Lu},
  \citenamefont {Kaminer}, \citenamefont {Pick}, \citenamefont {Chua},
  \citenamefont {Joannopoulos},\ and\ \citenamefont
  {Solja{\v{c}}i{\'c}}}]{zhen2015spawning}%
  \BibitemOpen
  \bibfield  {author} {\bibinfo {author} {\bibfnamefont {B.}~\bibnamefont
  {Zhen}}, \bibinfo {author} {\bibfnamefont {C.~W.}\ \bibnamefont {Hsu}},
  \bibinfo {author} {\bibfnamefont {Y.}~\bibnamefont {Igarashi}}, \bibinfo
  {author} {\bibfnamefont {L.}~\bibnamefont {Lu}}, \bibinfo {author}
  {\bibfnamefont {I.}~\bibnamefont {Kaminer}}, \bibinfo {author} {\bibfnamefont
  {A.}~\bibnamefont {Pick}}, \bibinfo {author} {\bibfnamefont {S.-L.}\
  \bibnamefont {Chua}}, \bibinfo {author} {\bibfnamefont {J.~D.}\ \bibnamefont
  {Joannopoulos}},\ and\ \bibinfo {author} {\bibfnamefont {M.}~\bibnamefont
  {Solja{\v{c}}i{\'c}}},\ }\bibfield  {title} {\bibinfo {title} {Spawning rings
  of exceptional points out of dirac cones},\ }\href@noop {} {\bibfield
  {journal} {\bibinfo  {journal} {Nature}\ }\textbf {\bibinfo {volume} {525}},\
  \bibinfo {pages} {354} (\bibinfo {year} {2015})}\BibitemShut {NoStop}%
\bibitem [{\citenamefont {Doppler}\ \emph {et~al.}(2016)\citenamefont
  {Doppler}, \citenamefont {Mailybaev}, \citenamefont {B{\"o}hm}, \citenamefont
  {Kuhl}, \citenamefont {Girschik}, \citenamefont {Libisch}, \citenamefont
  {Milburn}, \citenamefont {Rabl}, \citenamefont {Moiseyev},\ and\
  \citenamefont {Rotter}}]{doppler2016dynamically}%
  \BibitemOpen
  \bibfield  {author} {\bibinfo {author} {\bibfnamefont {J.}~\bibnamefont
  {Doppler}}, \bibinfo {author} {\bibfnamefont {A.~A.}\ \bibnamefont
  {Mailybaev}}, \bibinfo {author} {\bibfnamefont {J.}~\bibnamefont {B{\"o}hm}},
  \bibinfo {author} {\bibfnamefont {U.}~\bibnamefont {Kuhl}}, \bibinfo {author}
  {\bibfnamefont {A.}~\bibnamefont {Girschik}}, \bibinfo {author}
  {\bibfnamefont {F.}~\bibnamefont {Libisch}}, \bibinfo {author} {\bibfnamefont
  {T.~J.}\ \bibnamefont {Milburn}}, \bibinfo {author} {\bibfnamefont
  {P.}~\bibnamefont {Rabl}}, \bibinfo {author} {\bibfnamefont {N.}~\bibnamefont
  {Moiseyev}},\ and\ \bibinfo {author} {\bibfnamefont {S.}~\bibnamefont
  {Rotter}},\ }\bibfield  {title} {\bibinfo {title} {Dynamically encircling an
  exceptional point for asymmetric mode switching},\ }\href@noop {} {\bibfield
  {journal} {\bibinfo  {journal} {Nature}\ }\textbf {\bibinfo {volume} {537}},\
  \bibinfo {pages} {76} (\bibinfo {year} {2016})}\BibitemShut {NoStop}%
\bibitem [{\citenamefont {Zhou}\ \emph {et~al.}(2018)\citenamefont {Zhou},
  \citenamefont {Peng}, \citenamefont {Yoon}, \citenamefont {Hsu},
  \citenamefont {Nelson}, \citenamefont {Fu}, \citenamefont {Joannopoulos},
  \citenamefont {Soljačić},\ and\ \citenamefont
  {Zhen}}]{zhou_observation_2018}%
  \BibitemOpen
  \bibfield  {author} {\bibinfo {author} {\bibfnamefont {H.}~\bibnamefont
  {Zhou}}, \bibinfo {author} {\bibfnamefont {C.}~\bibnamefont {Peng}}, \bibinfo
  {author} {\bibfnamefont {Y.}~\bibnamefont {Yoon}}, \bibinfo {author}
  {\bibfnamefont {C.~W.}\ \bibnamefont {Hsu}}, \bibinfo {author} {\bibfnamefont
  {K.~A.}\ \bibnamefont {Nelson}}, \bibinfo {author} {\bibfnamefont
  {L.}~\bibnamefont {Fu}}, \bibinfo {author} {\bibfnamefont {J.~D.}\
  \bibnamefont {Joannopoulos}}, \bibinfo {author} {\bibfnamefont
  {M.}~\bibnamefont {Soljačić}},\ and\ \bibinfo {author} {\bibfnamefont
  {B.}~\bibnamefont {Zhen}},\ }\bibfield  {title} {\bibinfo {title}
  {Observation of bulk fermi arc and polarization half charge from paired
  exceptional points},\ }\href {https://doi.org/10.1126/science.aap9859}
  {\bibfield  {journal} {\bibinfo  {journal} {Science}\ }\textbf {\bibinfo
  {volume} {359}},\ \bibinfo {pages} {1009} (\bibinfo {year}
  {2018})}\BibitemShut {NoStop}%
\bibitem [{\citenamefont {Miri}\ and\ \citenamefont
  {Alu}(2019)}]{miri2019exceptional}%
  \BibitemOpen
  \bibfield  {author} {\bibinfo {author} {\bibfnamefont {M.-A.}\ \bibnamefont
  {Miri}}\ and\ \bibinfo {author} {\bibfnamefont {A.}~\bibnamefont {Alu}},\
  }\bibfield  {title} {\bibinfo {title} {Exceptional points in optics and
  photonics},\ }\href@noop {} {\bibfield  {journal} {\bibinfo  {journal}
  {Science}\ }\textbf {\bibinfo {volume} {363}} (\bibinfo {year}
  {2019})}\BibitemShut {NoStop}%
\bibitem [{\citenamefont {Hatsugai}(1993)}]{hatsugai1993chern}%
  \BibitemOpen
  \bibfield  {author} {\bibinfo {author} {\bibfnamefont {Y.}~\bibnamefont
  {Hatsugai}},\ }\bibfield  {title} {\bibinfo {title} {Chern number and edge
  states in the integer quantum hall effect},\ }\href@noop {} {\bibfield
  {journal} {\bibinfo  {journal} {Phys. Rev. Lett.}\ }\textbf {\bibinfo
  {volume} {71}},\ \bibinfo {pages} {3697} (\bibinfo {year}
  {1993})}\BibitemShut {NoStop}%
\end{thebibliography}%

\end{document}